\documentclass[11pt,a4paper]{article}
\usepackage[margin=2.3cm]{geometry}
\usepackage{amsmath,amssymb,graphicx,booktabs,array,microtype,xcolor,caption,float}
\usepackage{xurl}
\usepackage[hidelinks]{hyperref}
\usepackage[authoryear,round]{natbib}
\newcommand{\Msun}{M_\odot}

\title{Limited discrimination of neutron-decay dark matter under nuclear equation-of-state uncertainty in phenomenological neutron-star populations}
\author{Wasif Husain\\ \small School of Science, Technology and Engineering, University of Sunshine Coast, Adelaide 5000}
\date{}

\begin{document}
\maketitle

\begin{abstract}
We test discrimination of neutron-decay dark matter under nuclear equation-of-state (EoS) uncertainty in phenomenological neutron-star populations. The main sample comprises 236 EoSs balanced across four families after numerical causality and dark-fermion mass-stability screening. Whole-EoS-held-out classification using mass, radius and $f$ gives 63\% accuracy without measurement error and 36\% with illustrative errors (25\% baseline). Finite-catalogue Bayesian evidence classification combining stars sharing one EoS raises four-family accuracy from 33\% for one mass--radius observation to 42\% for three stars with radius, tidal and $f$-mode observations (partly driven by the self-bound family), while nuclear-versus-dark-matter discrimination remains consistent with chance (50\%; conditional EoS-cluster interval 46--55\%). At $1.4\,M_\odot$ the nuclear frequency spread is 7.35\% and the median paired dark-sector shift is 0.50\% (ratio 0.068), ranging from 0.33 to 1.65\% across interaction bins. The shifts closely follow mean-density scaling: the median radius-adjusted frequency residual is 0.069\%, against nuclear residual scatter of 1.73\%, with no demonstrated incremental nuclear-versus-dark-matter discrimination from $f$ beyond mass and radius under the tested conditions. In controlled observable-shift injections, three-star discrimination first exceeds chance (simultaneous 95\% lower bound) at a tested multiplier $a=3$ (about 1.5\% in $f$) for both precise and low-noise error sets with full references; repeated reference subsampling moves the low-noise crossing to $a=2$. Restricted-catalogue results are inconclusive after permutation and multiplicity controls. Results are conditional on the catalogue, illustrative likelihoods and effective one-fluid response; they do not define a physical detection threshold.
\end{abstract}
\section{Introduction}
Neutron stars probe dense matter at densities beyond nuclear saturation \citep{LP04,OF16}. Neutron decay into a dark sector has been proposed in connection with the neutron lifetime discrepancy \citep{FG18}. The existence of massive neutron stars strongly constrains scenarios with weakly interacting dark baryons \citep{McKeen18,Baym18,Motta18}; repulsive interactions can alleviate this constraint \citep{Husain22}. A change in stellar structure or oscillation frequency predicted for a specified nuclear equation of state (EoS) does not, however, establish that the dark component can be identified when that nuclear EoS is uncertain.

Degenerate macroscopic observables are well known in dense-matter physics \citep{Alford05}. Dark-matter effects on neutron-star $f$-mode observables have also been investigated \citep{Shirke24,HusainOsc}. In particular, our earlier published study considered selected QMC-based stellar models and gravitational-wave detectability \citep{HusainOsc}. Here the distinct question is the statistical discrimination of a specified neutron-decay dark sector across variable nuclear EoSs, including several stars sharing one EoS. We do not claim the underlying particle model or the existence of an $f$-mode shift as new. Machine learning has previously been used for EoS and composition inference \citep{Fujimoto18,Carvalho23,Carvalho24Hyp}; our classification labels denote generating families rather than microscopic species measurements. The test is complementary to continuous shared-EoS inference from multiple astrophysical observations \citep{LandryEssick19,Raaijmakers21,Miller21,Legred21}.

We compare matched nuclear and admixed configurations, evaluate classifiers on held-out whole EoSs, and calculate finite-catalogue Bayesian evidence for single-star and shared-EoS observations, including tidal deformability. The main sample satisfies both a numerical causality screen and a stable-dark-fermion mass restriction. The conclusions are conditional on these phenomenological families, the interaction prior, the effective-fluid response and the prescribed errors. An earlier fixed-sequence preprint benchmark is corrected in Appendix~\ref{sec:legacy}.
\section{A population of equations of state}\label{sec:pop}
\subsection{Composition scenarios as EoS populations}
Each scenario is a family of EoS defined by a small number of parameters with a uniform prior (Table~\ref{tab:priors}); one draw from the prior is one EoS. The families are phenomenological parametrisations \emph{inspired by} the four scenarios of the earlier version, not microphysical calculations. A composition label denotes the chosen generating family, \emph{not} a direct measurement of the species present at each density. In particular, the self-bound strange-matter family is \emph{not} a family of hybrid stars with a nuclear envelope and a quark core. The hyperonic-like family introduces a phenomenological softening and subsequent stiffening; this does not specify hyperon fractions as functions of density. A microscopic onset or a first-order phase transition would require a matched low- and high-density EoS, a transition pressure or chemical potential, and a choice of construction and interface response. Such transition locations are not inferred here.
\begin{table}[htbp]\centering\small
\caption{Parametrisations and priors. PP: piecewise polytrope \citep{Read09} with core break densities $\rho_1=10^{14.7}$ and $\rho_2=10^{15}$\,g\,cm$^{-3}$ on a four-piece crust; CSS: constant speed of sound, $\varepsilon=\varepsilon_0+p/c_s^2$.}\label{tab:priors}
\begin{tabular}{>{\raggedright\arraybackslash}p{.17\textwidth}p{.75\textwidth}}\toprule
Scenario & Parameters (uniform priors)\\\midrule
Nucleonic & PP: $\log_{10}P_1\in[34.15,34.60]$ (dyn\,cm$^{-2}$ at $\rho_1$), $\Gamma_1\in[2.4,3.8]$, $\Gamma_2\in[2.4,3.9]$, $\Gamma_3\in[2.2,3.9]$\\
Hyperonic-like & PP with $\log_{10}P_1\in[34.20,34.60]$, $\Gamma_1\in[2.6,3.8]$ and a soft segment $\Gamma_2\in[1.8,2.6]$ above $\rho_1\approx2n_0$ (phenomenological softening), then a causal CSS segment with $c_s^2\in[0.5,1]$ above $\rho_t=10^{u}$, $u\in[14.95,15.25]$\\
Strange (self-bound) & CSS with $\varepsilon_0=4B$, $B^{1/4}\in[130,160]$\,MeV, $c_s^2\in[1/3,1]$ ($c_s^2=1/3$ is the MIT bag model)\\
Dark-matter admixed & nucleonic PP prior plus a self-interacting dark Fermi gas in chemical equilibrium with the baryons (neutron-decay scenario): $m_\chi\in[937.9,939.2]$\,MeV, repulsive self-interaction $G\in[50,500]$\,fm$^2$; effective one-fluid barotrope\\\bottomrule\end{tabular}\end{table}
For the stable-dark-fermion benchmark, the relevant approximate mass interval is $937.9<m_\chi<938.783$\,MeV \citep{FG18}. The exploratory population-A prior extends to $939.2$\,MeV and is therefore \emph{not wholly physically admissible for that channel}. Out-of-window draws are treated solely as a sensitivity extension, not viable examples of the specified particle model. The joint-screened catalogue described below excludes these exploratory out-of-window draws.
Draws are accepted only if they satisfy broad observational and consistency screens: $M_{\max}\ge1.97\,\Msun$, a central sound speed satisfying $c_s^2\le1$ at the maximum-mass configuration, $70\le\Lambda_{1.4}\le800$ and $10\le R_{1.4}\le14$\,km. The acceptance rates are 21\%, 57\%, 24\% and 26\% for the nucleonic, strange, dark-matter and hyperonic-like priors, so the screens remove most of the prior volume (mainly through $M_{\max}$ and causality, and for strange matter through $\Lambda_{1.4}$ and $R_{1.4}$). Archived stellar mass grids vary by EoS; inference uses the common interval $1.2$--$1.85\,\Msun$, where all accepted EoS have stable stars. The initial screening tested causality at the maximum-mass centre only; a later full-interior audit uncovered nine cases with superluminal sound speed at intermediate pressures. The joint-screened sample is defined below.

The dark-matter-admixed scenario follows the effective-barotrope construction of the earlier work: baryons (nucleonic PP, $\mu_N=(\varepsilon+p)/n_B$) and a dark Fermi gas with $\mu_\chi=\sqrt{m_\chi^2+k_F^2}+\hbar cGn_\chi$ share one chemical potential, and the total pressure and energy density are summed at that potential. The implemented threshold is $n_\chi=0$ when $\mu_N\le m_\chi$; otherwise the positive chemical-equilibrium solution is used. The calculation adopts a calibrated low-density crust and a reference baryon mass. This is a model prescription, not a microscopic crust conversion calculation. The dark fermion is assigned baryon number $B_\chi=1$ in the baryon-conserving conversion benchmark. Writing $n_N=n_n+n_p$ for the nuclear baryon density, the conserved total baryon density is $n_{B,\mathrm{tot}}=n_N+n_\chi$; the quantity $n_B$ in $\mu_N=(\varepsilon_N+p_N)/n_B$ denotes the \emph{nuclear} density only. The conversion condition on the background is $\mu_N=\mu_\chi$; it does not itself determine the perturbative conversion timescale. By construction the DM scenario shares its nuclear prior with the nucleonic family. The classifier therefore tests whether the weak DM signature survives uncertainty in the baseline nuclear EoS.

\subsection{Solvers and validation}\label{sec:solver}
\paragraph{TOV and tidal deformability.} Stars are integrated with the Tolman--Oppenheimer--Volkoff equations \citep{Tolman39,OV39} and the tidal-deformability equation \citep{Hinderer08}.
\paragraph{Full-GR $f$-mode.} The $l=2$ quasi-normal $f$-mode of the non-rotating, barotropic star is computed from the linearised Einstein equations in the Regge--Wheeler gauge. The interior equations were derived symbolically (not taken from the literature) from the linearised Einstein equations and the conservation law, reduced to four first-order equations, and solved with regular boundary conditions at the centre and a vanishing Lagrangian pressure perturbation at the surface; this is the same physical system as the Lindblom--Detweiler formulation \citep{DL85}. The exterior solution is matched to the Zerilli equation \citep{Zerilli70}, and the complex eigenfrequency is the zero of the ingoing-wave amplitude; the frequency is $f=\operatorname{Re}\omega/(2\pi)$ and $\tau=1/\operatorname{Im}\omega$ in the stated time convention (with consistent physical units). The exterior equations were checked against an independent symbolic derivation (residual $10^{-12}$). The complex eigenfrequency is found by a secant search on the analytic ingoing amplitude, starting within 3\% of the root.
\paragraph{Numerical error budget.} Varying one numerical setting at a time (interior and exterior integration tolerances from $10^{-8}$ to $10^{-13}$, the TOV tolerance, the surface cut-off, the starting radius at the centre and the outer matching radius from 1500 to 8000\,km) changes $f$ by at most $3\times10^{-8}$ in relative terms and $\tau$ by at most $5\times10^{-7}$, for SLy and for SLy with $G=100$\,fm$^2$ at $1.4\,\Msun$. The frequency variation is more than five orders of magnitude below the exploratory median dark-matter frequency shift (0.56\%). The remaining sensitivity is physical: changing the crust pressure scale by $\pm10\%$ shifts $f$ of the nuclear EoS by 0.20\% but changes the dark-matter shift only from 1.313\% to 1.338\% (SLy, $G=100$\,fm$^2$), so the shift is insensitive to this particular crust-scale variation, not to all crust treatments.
\paragraph{What the oscillation observables measure.}
For a nonrotating equilibrium model the complex $l=2$ eigenfrequency depends on the equilibrium EoS \emph{and} the perturbative constitutive response; the real part is therefore not universally an EoS-only observable. With the chosen convention $e^{i\omega t}$, the calculated damping time is the \emph{gravitational-wave radiation} e-folding time, $\tau_{\rm GW}=1/\operatorname{Im}\omega$, for a stable mode with $\operatorname{Im}\omega>0$. The associated power-spectrum Lorentzian has a narrow-mode full width at half maximum $\Delta f_{\rm GW}\simeq 1/(\pi\tau_{\rm GW})$. For weak, approximately additive dissipation,
\begin{equation}
  \tau_{\rm eff}^{-1}\simeq \tau_{\rm GW}^{-1}+
  \tau_{\rm bulk}^{-1}+\tau_{\rm shear}^{-1}+\cdots .
\end{equation}
Bulk and shear viscosity, mutual friction, entrainment, superfluid degrees of freedom and chemical relaxation can alter damping and, in general, the mode spectrum. None of those transport coefficients is calculated here. Accordingly, the tabulated $\tau$ is \emph{not} a prediction of the full astrophysical damping time; the noisy inference below does not use $\tau$.
\paragraph{Damping times.} Direct independent solver benchmarks for the damping times of precisely these six stellar configurations were not available. The time dependence is $e^{i\omega t}$, so $\tau=1/\mathrm{Im}\,\omega$ and a damped mode has $\mathrm{Im}\,\omega>0$. The exterior machinery that produces $\mathrm{Im}\,\omega$ (Zerilli integration, projection onto asymptotic ingoing and outgoing solutions, complex root search) was benchmarked on a case with a published answer: with the horizon matched to purely ingoing waves it gives the fundamental $l=2$ quasi-normal mode of a Schwarzschild black hole, $M\omega=0.37366+0.08896\,i$, against $0.37367+0.08896\,i$ \citep{Leaver85}, i.e.\ relative errors of 3e-05 (real part) and 8e-06 (imaginary part). The interior contribution to $\tau$ lacks a like-for-like independent stellar benchmark. We include an idealized ablation below, not a forecast of its astrophysical utility.
\paragraph{External consistency check for stellar damping times.}
The computed damping times were compared, without parameter refitting, against the nonrotating-star empirical relation of \citet{Benhar04},
\begin{equation}
\tau_{\rm BFG}=\frac{R^4}{cM_{\rm g}^3}
\left(0.087-0.271\frac{M_{\rm g}}{R}\right)^{-1},
\qquad M_{\rm g}=\frac{G_{\rm N}M_\star}{c^2},
\end{equation}
with $R,M_{\rm g}$ in km and $c$ in km\,s$^{-1}$. For the six nuclear EoS at $1.4\,\Msun$ in Table~\ref{tab:valid}, the solver results differ by 6.0--9.6\% (median absolute deviation 8.7\%) from this independently published scaling relation. This is a consistency check of the damping-time magnitude and compactness dependence, \emph{not} an independent complex-eigenfrequency calculation for the same stars. The scientific interpretation of $\tau$ therefore remains conditional on the perturbation solver.

\paragraph{Independent compactness-fit check.} We additionally compare the archived frequencies and radiation damping times with the published hadronic and pure-quark compactness relations of \citet{ZhaoLattimer22}, Eqs.~(16)--(17) and Table~II, without refitting. The dimensionless complex frequency is $\Omega_f=G_{\rm N}M_\star\omega/c^3$. At $1.4\Msun$, the maximum absolute fractional deviations of the six nuclear configurations from the hadronic fit are 1.80\% for $f$ and 3.72\% for $\tau$; applying that fit to the twelve matched admixed configurations gives 1.77\% and 3.81\%. For all 90 causal CSS EoSs at $1.5\Msun$, the median, 95th percentile and maximum absolute deviations from the pure-quark fit are respectively 2.54, 4.46 and 5.08\% for $f$, and 2.86, 4.29 and 4.73\% for $\tau$ (log-PCHIP interpolation only if the mass is not tabulated). These checks corroborate the broad frequency and damping scales, including for self-bound stars. Empirical relations have intrinsic scatter, so they do not independently validate subpercent dark-sector shifts or the exact surface matching of the eigenmode solver.

\paragraph{Adiabatic response of the dark-matter configurations.} The perturbation equations use the equilibrium barotrope, i.e.\ chemical equilibrium is assumed to be maintained during the oscillation. As a restricted test of the opposite response prescription, $f$ was recomputed with a frozen-composition adiabatic response (baryon and dark numbers separately conserved, both components compressed together) for 24 dark-matter EoS at $1.4\,\Msun$: the median dark-matter shift is 0.64\% with the equilibrium response and 0.63\% with the frozen one, a difference of 1.1\% of the shift (at most 0.021\% of $f$). For the tested 24 configurations, the shift is insensitive at the reported level to the \emph{common-displacement} frozen-versus-equilibrium prescription. This is not a bound for genuine two-fluid oscillations: independent displacements, entrainment, chemical-relaxation timescales and nuclear composition gradients are not included. Hence the subpercent DM signature remains conditional on the chosen perturbative model.
\paragraph{Separate-fluid hydrostatic check.} As an additional background diagnostic, the nuclear and dark hydrostatic pressure gradients were integrated separately in a common spacetime metric for SLy, AP4 and MPA1 at $1.4\,\Msun$ with $G=100$ and $300\,\mathrm{fm}^2$. For these six configurations the maximum absolute radius difference from the effective one-fluid TOV construction was $0.211\%$; the total mass was matched to better than $3\times10^{-5}\%$. This tests the equilibrium stellar \emph{background} only, not the perturbative dynamics or the existence of distinct dark and nuclear normal modes. The numerical details and output are in \texttt{population/twofluid\_background.py} and \texttt{results\_twofluid\_background.json}.\par
\paragraph{Validation.} The crust is a single calibrated parameter (the absolute pressure scale of a four-piece SLy-like crust, 1.135 times the electron-gas value), fitted only to the radius and tidal deformability of six published EoS fits at $1.4\,\Msun$ (Table~\ref{tab:valid}); the radii are then reproduced to 0.0006\,km and $\Lambda$ to 0.05\%. The $f$-mode frequencies were \emph{not} used in any fit, and are reproduced to 0.0072\% (maximum over the six EoSs). The reference values are those of the finite-EoS pilot of the earlier work (\citealt{HusainPilot}), which used the separately developed RelModPy full-GR code (\url{https://github.com/fgittins/RelModPy}); the six values are listed in Table~\ref{tab:valid} so that the comparison can be repeated with any other code. The pilot manuscript is unpublished and its standalone text was not recovered for this submission; the archived reference values and comparison script are supplied with the electronic supplementary material. This is a same-author comparison with a separately developed code, not an independently reproduced published benchmark. In addition, the dark-matter frequency shift at fixed nuclear EoS ($\mathrm{SLy}$, $G=100$\,fm$^2$, $1.4\,\Msun$) is 25.7\,Hz against 25.8\,Hz in the reference, and a tabulated version of the same EoS agrees with the analytic one to 0.013\% in $f$. Damping times of $0.17$--$0.23$\,s are obtained for the six EoS, as expected for a $1.4\,\Msun$ $f$-mode; no independent reference values for $\tau$ were available.
\begin{table}[htbp]\centering\small
\caption{Same-author pilot code comparison at $1.4\,\Msun$ using six published nuclear EoS fits. $R$ and $\Lambda$ calibrated the crust; $f$ was not fitted. The reference frequencies are unpublished.}\label{tab:valid}
\begin{tabular}{lrrrrrrrr}\toprule
EoS & $R$ (km) & $R_{\rm ref}$ & $\Lambda$ & $\Lambda_{\rm ref}$ & $f$ (Hz) & $f_{\rm ref}$ & $\Delta f/f$ (\%) & $\tau$ (s)\\\midrule
SLy & 11.7055 & 11.7061 & 295.5 & 295.6 & 1935.20 & 1935.06 & +0.0072 & 0.1920 \\
AP3 & 11.9712 & 11.9717 & 367.1 & 367.2 & 1848.51 & 1848.40 & +0.0059 & 0.2094 \\
AP4 & 11.3192 & 11.3197 & 247.5 & 247.6 & 2007.63 & 2007.50 & +0.0062 & 0.1793 \\
ENG & 11.9695 & 11.9699 & 369.9 & 370.1 & 1845.65 & 1845.53 & +0.0066 & 0.2100 \\
MPA1 & 12.4339 & 12.4343 & 478.9 & 479.1 & 1745.68 & 1745.58 & +0.0057 & 0.2341 \\
WFF2 & 11.1245 & 11.1249 & 224.9 & 225.0 & 2047.47 & 2047.34 & +0.0062 & 0.1729\\\bottomrule\end{tabular}\end{table}

\subsection{Numerical and particle-stability selection}
From 90 initially centre-screened EoSs per family, the full-interior audit retains 85 nucleonic, 90 self-bound, 86 admixed and 90 hyperonic-like EoSs. The test requires $0<c_s^2\le 1+10^{-3}$ at 256 pressure points from the surface to the maximum-mass central pressure; it is a numerical test rather than an analytic guarantee. The dark-fermion cut $937.9<m_\chi\le938.783$ MeV retains 59 of the 86 causal admixed models. We take the first 59 eligible indices in each class (236 total), using equal family priors. The wider causal-only catalogue of 340 EoSs (85 per class) is a robustness check, not the physical dark-mass benchmark. No acausal EoS is included in either classifier or Bayesian evidence evaluation. The two catalogues overlap and are not independent experiments.

The original screen allows massive self-bound models; it does not adopt a hard upper mass bound based on merger-remnant assumptions. To expose selection sensitivity, Table~\ref{tab:selection} reports the conditional retention by interaction bin, together with the counts that would survive an illustrative upper maximum-mass cut of $2.3\,\Msun$. These are conditional counts, not original proposal acceptance rates or a new constraint on $G$; restricted inference is reported separately in Sec.~\ref{sec:sensitivity}.
\begin{table}[htbp]\centering\small
\caption{Conditional retention of the original 90 accepted admixed EoSs (left). Right: counts of causality-screened EoSs before and after an illustrative maximum-mass ceiling. Counts are conditional on the accepted catalogue, not proposal-level acceptance probabilities.}\label{tab:selection}
\begin{tabular}{lrrr@{\hspace{1.2em}}lrr}\toprule
$G$ (fm$^2$)& Initial & Causal & Joint & Family & Causal & $M_{\max}\le2.3\Msun$\\\midrule
50--100 &13&12&9&Nucleonic&85&64\\
100--200&19&19&12&Self-bound&90&20\\
200--300&20&19&11&Admixed&86&75\\
300--400&21&19&16&Hyperonic-like&90&89\\
400--500&17&17&11&&&\\\bottomrule
\end{tabular}\end{table}
\section{Inference methods}
\subsection{Single-star classification}
We split the EoSs within each class into five folds, retaining one for testing, one for calibration and three for training, and repeat with three seeded partitions. No EoS occurs in both training and test within a fold. The observables are interpolated on a common mass interval $1.2$--$1.85\,\Msun$. The primary noise-free and noisy feature vectors are $(M,R,f)$; an additional noise-free ablation includes the computed radiation damping time $\tau$. The noisy vector uses fractional standard deviations $(0.10,0.09,0.20)$. These illustrative errors are motivated by NICER mass--radius measurements \citep{Riley19,Miller19,Choudhury24,Salmi24} and prospective inspiral $f$-mode measurements \citep{Williams22}, not a joint observation of a real star.

For each EoS, 33 equally spaced masses are used with one draw per mass in the noise-free case or three noise draws otherwise. The MLP ensemble \citep{Lakshminarayanan17} is implemented in scikit-learn \citep{Pedregosa11} and comprises three networks, two hidden layers of 64 ReLU units, Adam with learning rate $10^{-3}$, L2 penalty $10^{-4}$, batch size 128 and up to 120 iterations with early stopping. A logistic-regression baseline uses the same log-transformed, training-standardized features. These are fixed settings in the screened experiments, not a claim of optimized performance. For the revised main classifier table we pool confusion counts within each held-out EoS over repetitions and bootstrap EoSs within class (2,000 resamples); reported intervals are conditional on the fitted models and catalogue, not independent partition confidence intervals. The four-class confusion statistic restricted to true and predicted labels both in the nuclear/DM block is distinguished from the independent binary evidence experiment below.

\subsection{Shared-EoS Bayesian evidence}\label{sec:evidence}
Let $D_i$ denote the observables for star $i$ and $E_e$ a reference EoS. Conditional on $E_e$, we draw stellar masses independently from a uniform prior on $[1.2,1.85]\,\Msun$. With independent fractional Gaussian errors truncated to positive values, the marginal likelihood for each star is
\begin{equation}
 L_i(E_e)=\int_{1.2\Msun}^{1.85\Msun}p(D_i\mid M_i,E_e)\,\pi(M_i)\,dM_i.
\end{equation}
The density for a positive scalar observation $x$ with model value $\mu$ and fractional error $\sigma$ is $\phi[(x-\mu)/(\sigma\mu)]/[\sigma\mu\Phi(1/\sigma)]$, where $\phi$ and $\Phi$ are the standard normal density and cumulative distribution. Observables within a star are treated as independent in this illustrative experiment. Each stellar mass is separately integrated, but all stars in a synthetic system share the same EoS. The finite-reference evidence for family $k$ is
\begin{equation}
 Z_k(D)=\frac{1}{n_k}\sum_{e\in k}\prod_i L_i(E_e),\qquad
 p(k\mid D)=\frac{Z_k}{\sum_j Z_j},\qquad B_{\rm DM,N}=Z_{\rm DM}/Z_{\rm N}.
\end{equation}
The product is formed before averaging over EoSs; multiplying already marginalized family probabilities would instead allow different EoSs for different stars. This approximation is Bayes evidence for the discrete reference catalogue, not for a fully integrated continuous physical prior.

We use stratified five-fold EoS holdout, with all four non-test folds as the reference; no trained classifier or calibration fold is needed. A log-PCHIP interpolant on the common mass interval defines the reference observables. Uniform midpoint quadrature uses 66, 132 and 264 points for the joint-screened catalogue and 132 and 264 for the causal-only catalogue. For each held-out EoS, eight independent synthetic systems are drawn in each of three seeded partitions. All stars in each system share that EoS. Five scenarios are listed in Table~\ref{tab:scenarios}. The binary experiment restricts the true and reference classes to nuclear and admixed EoSs and renormalizes their probabilities, rather than selecting predictions from a four-class confusion matrix.

\begin{table}[htbp]\centering\small
\caption{Illustrative shared-EoS observing scenarios. Each tuple lists a separate star and its fractional standard deviations in percent. $\Lambda$ is dimensionless tidal deformability. These are not actual NICER or gravitational-wave likelihoods.}\label{tab:scenarios}
\begin{tabular}{lp{.64\textwidth}}\toprule
Scenario & Observables (fractional errors)\\\midrule
One MR & $(M,R):(10,9)$\\
One MRf & $(M,R,f):(10,9,20)$\\
Two stars & $(M,R):(10,9)$; $(M,\Lambda):(5,30)$\\
Three stars & $(M,R):(10,9)$; $(M,\Lambda):(5,30)$; $(M,f):(5,20)$\\
Precise three stars & $(M,R):(3,3)$; $(M,\Lambda):(3,10)$; $(M,f):(3,0.5)$\\\bottomrule
\end{tabular}\end{table}

Following the logarithmic scoring rule \citep{GneitingRaftery07}, for equal class priors the held-out predictive information score is $S=\ln K+\langle\ln \widehat p(Y\mid D)\rangle$ in nats, where $K$ is the number of classes. It is an out-of-sample log-score improvement over the uniform predictor, not the exact mutual information of an unknown continuous population. Negative values are possible. We report 2,000 stratified EoS-cluster bootstrap percentile intervals for accuracy and the score: all repeated-partition and noise observations belonging to the same EoS are resampled together. The intervals are conditional on the fixed catalogue and reference folds; they do not include population-prior uncertainty or refitting of reference catalogues. Class precision and recall are computed from the EoS-weighted confusion matrix.
\section{Results}
\subsection{Screened single-star classifications}
Table~\ref{tab:ml} gives the rerun of the 236-EoS main catalogue with EoS-cluster intervals. With $M,R,f$, the noise-free MLP accuracy is 63.08\% and the noisy accuracy is 36.46\%; adding $\tau$ to the noise-free vector gives 63.15\%, a negligible observed change in this experiment. The conditional nuclear--DM statistic is calculated from the four-family confusion block and is not independently normalized binary inference. The earlier wider causal-only comparisons used $\tau$ in the noise-free vector (63.1\% MLP and 61.4\% logistic), and MRf with noise (36.7\% and 35.9\%); they are not the primary physical sample.
\begin{table}[htbp]\centering\small
\caption{Joint-screened sample (236 EoSs). Accuracy and conditional nuclear--DM confusion statistic (\%), with 95\% stratified EoS-cluster percentile intervals in brackets. Predictions are pooled within each EoS across repetitions.}\label{tab:ml}
\begin{tabular}{llrr}\toprule
Scenario & Classifier & Accuracy [95\%] & Conditional N--DM [95\%]\\\midrule
Noise-free MRf & MLP & 63.08 [59.39, 66.94] & 50.84 [43.58, 57.79] \\
Noise-free MRf & Logistic & 61.25 [57.44, 65.03] & 51.03 [44.75, 57.63] \\
Noise-free MRf$\tau$ & MLP & 63.15 [59.14, 66.99] & 50.34 [42.95, 57.54] \\
Noise-free MRf$\tau$ & Logistic & 61.74 [58.00, 65.68] & 51.27 [44.85, 57.88] \\
Noisy MRf & MLP & 36.46 [35.03, 37.96] & 49.40 [46.93, 52.15] \\
Noisy MRf & Logistic & 35.24 [33.64, 36.80] & 48.28 [44.20, 52.21]\\\bottomrule
\end{tabular}\end{table}
\subsection{Shared-EoS evidence and tidal observations}
\begin{table}[htbp]\centering\small
\caption{Joint-screened catalogue: held-out shared-EoS evidence with 264 mass quadrature points. Accuracy in percent, with EoS-cluster bootstrap percentile intervals in brackets. The information scores $S_4,S_2$ are in nats. Four-class and binary baselines are 25\% and 50\%.}\label{tab:evidence}
\resizebox{\textwidth}{!}{\begin{tabular}{lrrrr}\toprule
Scenario & Four-class accuracy [95\%] & Nuclear--DM [95\%] & $S_4$ & $S_2$\\\midrule
One MR & 32.52 [30.81, 34.16] & 48.73 [45.30, 52.40] & 0.060 & -0.003\\
One MRf & 35.88 [34.06, 37.75] & 49.33 [45.76, 53.11] & 0.088 & -0.003\\
Two stars & 41.61 [39.41, 43.86] & 50.60 [46.65, 54.84] & 0.178 & -0.005\\
Three stars & 42.41 [40.15, 44.62] & 50.49 [46.43, 54.77] & 0.194 & -0.005\\
Precise three stars & 59.59 [56.50, 62.38] & 48.59 [43.71, 53.53] & 0.590 & -0.104
\\\bottomrule
\end{tabular}}\end{table}
The additional tidal observation and a third star with an $f$-mode measurement improve overall discrimination from 32.52\% (one MR star) to 42.41\% (three stars), while the high-precision scenario reaches 59.59\% (Table~\ref{tab:evidence}). The independently normalized nuclear--DM binary test gives 50.49\% [46.43,54.77] for the three-star scenario and 48.59\% [43.71,53.53] for the precise variant. The latter has a negative predictive information score: overconfident finite-reference predictions can perform worse in log score than the uniform binary predictor. This does not establish a negative physical mutual information or a universal impossibility of detecting dark matter. For the wider 340-EoS causal-only comparison, the corresponding three-star and precise four-class accuracies are 42.49\% and 60.21\%; binary accuracies are 51.13\% and 49.98\%. The single-star binary accuracies for MR and MRf are 48.73\% and 49.33\%, respectively, with overlapping cluster intervals and nearly identical log scores ($-0.003$ nats). The added $f$ measurement does not demonstrate an incremental binary gain under these illustrative errors, although other families benefit.
For the joint-screened three-star case, class precision (recall) for the nucleonic, self-bound, admixed and hyperonic-like families is 32.9\% (38.7\%), 52.5\% (73.0\%), 28.9\% (16.3\%), 47.8\% (41.7\%). These values use equal class priors. The mean log Bayes factor $\ln B_{\rm DM,N}$ is -0.002 for true nuclear EoSs and -0.010 for true admixed EoSs. The reported joint-screened accuracies are unchanged to the quoted precision when the mass quadrature is increased from 66 or 132 to 264 points. This is a quadrature check only, not convergence with the size of the EoS catalogue.

\subsection{Matched dark-sector shifts and interaction dependence}
\begin{figure}[htbp]\centering\includegraphics[width=.92\textwidth]{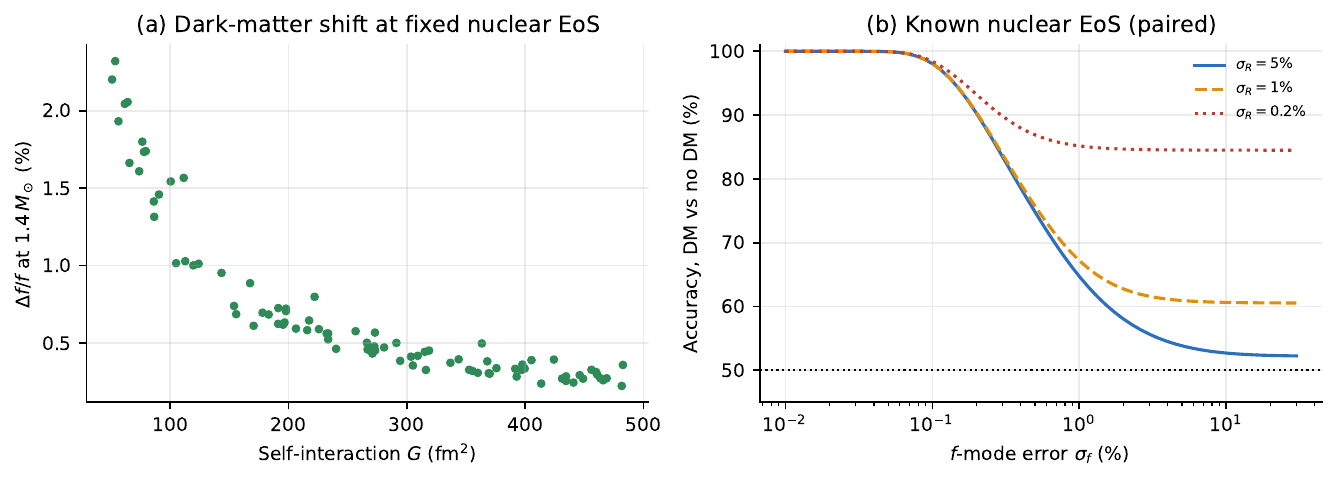}
\caption{Exploratory original paired catalogue (before the joint screen): frequency shifts and illustrative conditional discrimination with a known nuclear baseline. The scatter and forecast are not results on the main 236-EoS sample.}\label{fig:paired}\end{figure}

A separate exploratory paired calculation on the original 90 admixed EoSs and their own nuclear bases gives a median frequency increase of 0.56\% (5th--95th percentiles 0.27--2.18\%) at fixed mass, and a median radius change of $-0.40\%$. These descriptive results precede the joint screen; they are not estimates for the 236-EoS catalogue. The correlation of $\ln G$ with $\Delta f/f$ at $1.4\,\Msun$ is $-0.95$, illustrating the strong dependence on the assumed interaction. Figure~\ref{fig:paired} shows the exploratory paired shifts and an illustrative equal-covariance Gaussian discrimination calculation conditional on knowing the nuclear baseline exactly. The latter is not the evidence calculation of Sec.~\ref{sec:evidence}.

An additional matched calculation for six published nuclear PP representations (SLy, AP3, AP4, ENG, MPA1 and WFF2), at $1.4\,\Msun$ and $m_\chi=938$ MeV, gives median frequency increases of 1.215\% at $G=100$ fm$^2$ (range 1.144--1.326\%) and 0.400\% at $G=300$ fm$^2$ (range 0.380--0.442\%). The nuclear frequencies span 1746--2047 Hz; all twelve admixed configurations meet the lower maximum-mass criterion of $1.97\,\Msun$. The larger shifts at $G=100$ fm$^2$ relative to the 0.56\% exploratory median are expected because the latter averages over the broad 50--500 fm$^2$ prior. These calculations are nuclear-baseline checks, not independent oscillation benchmarks. In the interaction-dominated limit, writing $\delta\mu=\mu_N-m_\chi>0$, chemical equilibrium gives $n_\chi\simeq\delta\mu/(\hbar cG)$ if the dark kinetic contribution to $\delta\mu$ is small. The associated vector pressure $P_{\rm int}\simeq\delta\mu^2/(2\hbar cG)$ provides a qualitative explanation for the decreasing perturbation as $G$ increases; it is not an exact stellar frequency scaling. The conditional retention in Table~\ref{tab:selection} documents how screening changes the accepted interaction distribution. No inference is made outside the sampled interaction interval.

\subsection{Restricted-catalogue controls and signal relative to nuclear spread}\label{sec:sensitivity}
We repeat the three-star calculation with $M_{\max}\leq2.3\Msun$, interaction bins (condition DM only) and central rectangular nuclear parameter restrictions $\max_j|2(\theta_j-L_j)/(U_j-L_j)-1|\le w$, with the bounds in Table~\ref{tab:priors}. These are empirical restrictions, not posterior constraints. The mass cut is illustrative, not a universal physical upper bound. To avoid differences due to re-seeding, all restrictions inherit the original full-catalogue fold assignments and noise seeds (eight systems per EoS, three partitions, 132 quadrature points). For each fold, we also sample the larger reference classes without replacement down to the smallest reference count, 20 times, while retaining all held-out EoSs and equal family priors. We average prediction metrics across the subsamples, not their posterior probabilities. Conditional cluster intervals resample test EoSs, not refitted catalogues. Equal-reference controls probe count effects but cannot establish adequate sampling of either prior.
\begin{table}[htbp]\centering\scriptsize
\caption{Count-matched three-star binary accuracy (\%, conditional cluster interval), central 95\% label-permutation null range, nominal lower-tail permutation probability, and fixed-mass signal diagnostics. Nuclear spread $\sigma^N_f$ includes a 95\% percentile interval; paired median shifts $s_f,s_R,s_\Lambda$ are percentages at $1.4\Msun$. All pairs in each DM selection have a matched nuclear baseline.}\label{tab:sensitivity}
\resizebox{\textwidth}{!}{\begin{tabular}{llrrrrrrrr}\toprule
Selection & N/DM & Equal references & Null range & $p_-$ & $\sigma^N_f$ & $s_f$ & $|s_f|/\sigma^N_f$ & $s_R$ & $s_\Lambda$\\\midrule
Full prior & 59/59 & 50.49 [46.43,54.77] & [44.88,55.34] & 0.575 & 7.35 [5.93,8.30] & 0.50 & 0.068 & -0.35 & -2.30 \\
$M_{\max}\leq2.3\Msun$ & 43/53 & 48.15 [44.92,51.54] & [45.36,55.86] & 0.305 & 5.62 [4.11,7.57] & 0.46 & 0.082 & -0.33 & -2.25 \\
$50\le G<150$ & 59/14 & 44.90 [41.88,48.04] & [44.13,55.93] & 0.045 & 7.35 [5.86,8.37] & 1.65 & 0.224 & -1.15 & -7.23 \\
$150\le G<300$ & 59/18 & 46.74 [44.06,49.54] & [45.27,57.29] & 0.115 & 7.35 [5.91,8.30] & 0.59 & 0.080 & -0.41 & -2.76 \\
$300\le G\le500$ & 59/27 & 53.12 [47.57,58.51] & [45.68,57.11] & 0.850 & 7.35 [5.91,8.33] & 0.33 & 0.045 & -0.23 & -1.64 \\
$w=0.95$ & 44/49 & 47.57 [44.18,51.34] & [45.25,57.20] & 0.215 & 6.78 [4.87,7.84] & 0.50 & 0.074 & -0.35 & -2.37 \\
$w=0.85$ & 32/28 & 42.33 [38.93,45.86] & [43.83,57.42] & 0.015 & 7.13 [4.20,8.10] & 0.49 & 0.069 & -0.35 & -2.32 \\
$w=0.75$ & 23/21 & 44.11 [39.90,48.70] & [43.58,58.84] & 0.065 & 4.87 [2.92,6.68] & 0.48 & 0.098 & -0.34 & -2.27 \\
$w=0.65$ & 14/13 & 44.86 [41.06,49.15] & [43.55,58.42] & 0.075 & 3.36 [1.37,5.90] & 0.41 & 0.123 & -0.29 & -2.04 \\
\\\bottomrule\end{tabular}}\end{table}
The unrestricted control recovers 50.49\% [46.43,54.77]. Equalizing reference counts does not remove the low-$G$ below-chance result. The conditional label-permutation controls and dependence-robust joint tests (Bonferroni $p=0.135$; harmonic-corrected $p=0.382$) do not establish systematic below-chance discrimination. The mass cut retains only ten self-bound EoSs; the three-family calculation excluding them still shows a multi-star gain. These restricted-catalogue results are inconclusive rather than evidence for or against a physical discrimination gain. Full control details are in Appendix~\ref{sec:controls}.

At fixed mass define the nuclear frequency spread $\sigma^N_f=(Q_{84}(f_N)-Q_{16}(f_N))/(2\,\mathrm{median}(f_N))$ and the paired shifts $s_X=\mathrm{median}[(X_{\rm DM}-X_{\rm base})/X_{\rm base}]$ for $X=f,R,\Lambda$. Here $X_{\rm base}$ is evaluated for the same nuclear parameters as the admixed configuration; the nuclear spread is measured across the separately selected nuclear reference family. Table~\ref{tab:sensitivity} uses 59 matched pairs for the full screened sample, not the earlier 90-pair exploratory statistic. The full-catalogue $f$ spread is 7.35\% [5.93,8.30] and the median shift is 0.50\% [0.39,0.59], so $|s_f|/\sigma^N_f=0.068$ [0.052,0.092] (conditional 95\% percentile intervals from separately resampling nuclear EoSs and matched pairs). Thus the characteristic residual nuclear frequency uncertainty would have to be of order 0.5\% at fixed mass for the median shift to be comparable to the residual spread, a roughly fifteen-fold reduction relative to this catalogue. In the lowest $G$ bin the paired shift is larger (1.65\%), but the ratio is only 0.224. The narrowest empirical nuclear restriction still gives 3.36\% spread and 0.41\% signal. The apparent increase in spread from $w=0.95$ to $0.85$ is consistent with overlapping bootstrap intervals; monotonic narrowing is not established. The paired median tidal-deformability shift is $-2.30\%$ for the full sample and $-7.23\%$ in the lowest-$G$ bin. Its larger fractional magnitude is consistent with the strong compactness dependence of $\Lambda=(2/3)k_2(R/M)^5$ in geometric units, but changes in $k_2$ also matter; this is not an independent detection criterion. A rectangular parameter cut is not an independently measured nuclear EoS. This comparison measures signal size versus population spread, not the required measurement precision for a specified Bayes factor, nor a sufficient detection criterion. Observational noise, mass uncertainty, correlations and theoretical systematics are additional factors.
\subsection{Structural frequency residual and controlled signal injection}
At fixed mass, the 59 screened paired configurations give a median $100\Delta\ln f=0.496\%$ and $100\Delta\ln R=-0.350\%$, with Pearson correlation $-0.9996$. For the separate 59 nuclear EoSs at $1.4\Msun$, an ordinary least-squares fit $\ln f=a+b\ln(R/\mathrm{km})$ gives $b=-1.215$ ($R^2=0.947$). The paired median residual $100(\Delta\ln f-b\Delta\ln R)$ is 0.069\% [0.037,0.104], while the robust half 16--84 percentile range of leave-one-EoS-out nuclear log-frequency residuals is 1.73\% [1.19,2.08]. The latter interval resamples the fixed leave-one-out residuals; the paired interval resamples matched pairs and refits the nuclear slope on bootstrap draws. Under the approximate mean-density law $f\propto(M/R^3)^{1/2}$, the paired median $100(\Delta\ln f+1.5\Delta\ln R)$ is $-0.027\%$. These results indicate that the specified dark component largely follows the structural $f$--$R$ trend, rather than producing a separately resolved frequency signature. Residuals are smaller than the independent compactness-fit deviations, so this is a descriptive property of the model output, not a validated subpercent composition diagnostic.

To isolate the effect of signal size, we construct counterfactual observables from the 59 screened matched pairs. For $X\in\{R,f,\Lambda\}$ at each mass, define $\ln X_a=\ln X_N+a(\ln X_{\rm DM}-\ln X_N)$ with $a=0,0.5,1,2,5,10,20,40$. Stellar mass is unchanged. At $a=3$, the median injected shifts at $1.4\Msun$ are $+1.500\%$ in $f$ and $-1.044\%$ in $R$, calculated as $100[\exp(a\,\mathrm{median}(\Delta\ln X))-1]$ from the unrounded paired log shifts; they are not three times the rounded fractional shifts. At $a=1$ the original paired models are recovered; $a=0$ makes the families identical and returns exactly 50\% balanced accuracy (ties receive half credit). Both members of a pair remain in the same held-out fold, so a paired baseline cannot leak into the training reference. We use 132 mass quadrature points, three five-fold partitions, eight systems per pair member, and common mass and noise draws across pair members and multipliers. The intervals resample whole pairs. The matched-pair population differs from the independent nuclear reference family in Table~\ref{tab:evidence}; this deliberately isolates signal size rather than retesting the original catalogue. For $a\ne0,1$ these are observable-space injections, not self-consistent EoSs or new values of $G$; the large multipliers 20 and 40 deliberately stress the estimator beyond physical validation.

For the physical paired signal ($a=1$), one-star MR and MRf accuracy are 51.3\% and 51.2\%; three-star accuracy is 51.7\% and the precise variant gives 49.1\%. At $a=10$ the corresponding accuracies are 59.8\%, 61.5\%, 64.3\% and 68.0\%. Using a descriptive 75\% balanced-accuracy target on the tested grid, the precise three-star scenario first exceeds it at $a=20$ (79.3\% [75.2,83.0]); the ordinary three-star scenario first exceeds it at $a=40$ (86.2\% [83.6,88.7]), as do the two one-star scenarios. At $a=20$ the ordinary three-star estimate is 73.8\% [70.8,76.9], so the location of the crossing is uncertain. This is an empirical operating point under the specified noise and matched-reference protocol, not a detection threshold for real stars, nor a map from signal multiplier to dark abundance.
We additionally reduce fractional errors to 1\% in mass and radius, 5\% in tidal deformability and 0.1\% in frequency, retaining the same three-star protocol. Both the original precise variant and this low-noise variant are evaluated on the extended grid $a=0,0.25,0.5,1,2,3,4,5,10,20,40$. We use 50,000 common pair-bootstrap resamples and report pointwise 95\% percentile intervals and Bonferroni simultaneous one-sided 95\% lower bounds across the 20 nonzero scenario--multiplier combinations (lower quantile $0.05/20$). These are conditional on the fixed references, not unconditional population confidence bounds. In the low-noise variant $a=1$ gives 49.65\% [47.21,52.05]; $a=2$ gives 54.84\% [51.48,58.16] (pointwise above chance), but its simultaneous lower bound is 49.93\%. The first tested point exceeding chance by the simultaneous criterion is $a=3$ (60.20\%, lower bound 55.23\%); the original precise variant also first crosses at $a=3$ (55.40\%, lower bound 50.46\%). The low-noise curve uses 528 mass quadrature points; changing from 132 to 528 changes its grid accuracies by at most 0.071 percentage points without changing either crossing. These are grid-dependent conditional sensitivity results, not a claim of detectability for actual stars or a precise continuous threshold. Non-monotonicity of the finite-reference curve is retained rather than smoothed away. The physical signal remains unresolved in the low-noise experiment.

A reference-density control retains 50, 75 or 100\% of the training pairs within each fold (approximately 24, 35--36 or 47--48 pairs), with both members retained together. The smaller sets are nested within 20 independently randomized orderings per fold; we average accuracies across these subsets, retaining the same test draws, 132-point quadrature and pair-bootstrap procedure. At each fraction the simultaneous bounds cover the same 20 nonzero scenario--multiplier combinations, not the three fractions jointly. The full-reference control reproduces the archived 132-point curve exactly to numerical precision. The precise variant first exceeds chance at $a=3$ for all fractions (lower bounds 50.22, 50.03 and 50.46\%); the low-noise variant first crosses at $a=2$ for the 50 and 75\% fractions (50.34 and 50.33\%), versus $a=3$ at full references. The unamplified signal remains consistent with chance in all cases. The near-boundary crossing is thus sensitive to the finite-reference estimator and subset-averaging protocol; the latter also changes conditional variance, so this is not evidence that reducing the reference size improves physical discrimination. Neither a continuous-prior limit nor the relative contributions of nuclear spread and estimator density are established.
\begin{figure}[htbp]\centering\includegraphics[width=.97\textwidth]{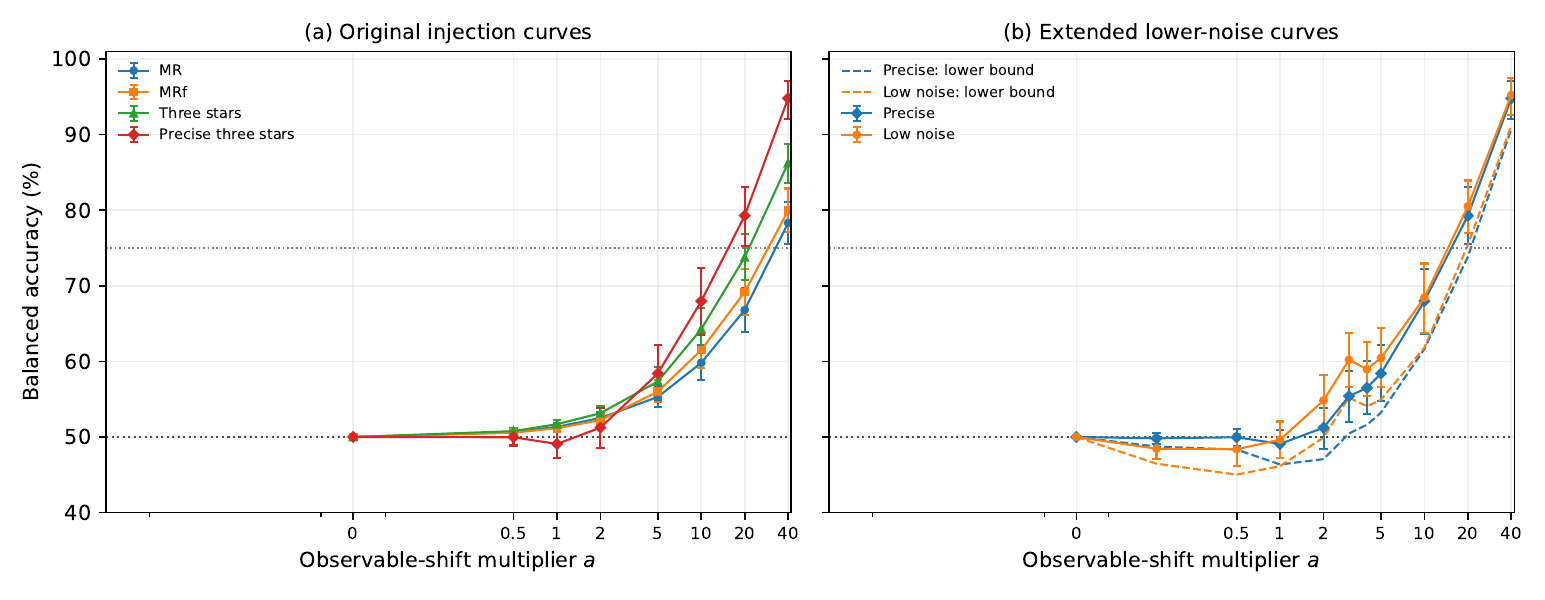}
\caption{Controlled paired-observable injection. The multiplier scales log shifts in radius, frequency and tidal deformability coherently at fixed mass. Panel (a): original four observation scenarios with conditional 95\% pair-bootstrap intervals. Panel (b): extended precise and low-noise three-star curves with pointwise intervals and simultaneous one-sided 95\% lower bounds (dashed). Horizontal lines indicate chance and the descriptive 75\% target. The points do not represent a sequence of physical EoSs.}\label{fig:injection}\end{figure}

\section{Discussion and conclusions}
The main result is a conditional one: within these jointly screened phenomenological populations, a finite-catalogue evidence calculation gains family-level discrimination by combining several stars, but the nuclear--DM binary predictions remain consistent with chance. The overlap is expected because the two families share the same nuclear prior and the adopted repulsive interaction yields modest changes to the barotrope. The quantitative result is useful as a stress test of a specified scenario, not as a discovery of a composition-independent physical degeneracy. Unlike the earlier fixed-sequence benchmark, it evaluates completely held-out EoSs and independent binary probabilities.

The assumed interaction prior, upper-mass selection, and phenomenological family definitions matter. A self-bound CSS family is not a generic quark-core population; softening and stiffening alone do not determine hyperon fractions. Neither the pressure-curve representational equivalence of nuclear and admixed models nor the discrimination achievable under a different $G$ prior is established. The restricted-catalogue reruns are inconclusive because of small, changing reference sets. The fixed-mass signal-to-spread and frequency--radius residual results, rather than small differences between restricted accuracies, quantify the competition between nuclear uncertainty and the paired dark-sector shift. The observable injection provides a complementary conditional sensitivity curve. We do not estimate dark abundance, impose real correlated multi-instrument likelihoods, or infer microscopic composition. Equal class probabilities and uniform stellar masses are study assumptions, not astrophysical population fractions. The finite reference may poorly approximate a continuous family, particularly at high measurement precision.

The frequency calculation uses an effective one-fluid equilibrium response. This reduction to a barotrope removes independent dark and baryonic displacement degrees of freedom, making the observed structural $f$--$R$ trend physically plausible (no universal relation is imposed on the solver). Independent dark and baryonic motion in a genuine two-fluid mode could break this behaviour and is not tested here. The separately integrated backgrounds and common-displacement frozen-response check do not validate such modes. The same-author pilot comparison checks implementation consistency; published compactness fits provide an independent consistency check, including self-bound models, but a published like-for-like stellar complex-eigenfrequency benchmark for these particular models is not available. The noisy evidence excludes $\tau$; the noise-free ablation is not a claim about its utility with real damping measurements. The earlier published detectability forecast is not revalidated by the redshift audit or by the present phenomenological models. The preprint of the earlier published study (\citealt{HusainOsc}; arXiv:2505.17355) reports that repulsively admixed QMC models approach the nuclear EoS and follow similar compactness scalings; this is qualitatively compatible with the small matched shifts found here. Its numerical pressure--energy tables and exact oscillation implementation were not recovered in a form enabling a matched eigenvalue calculation. Published curves and a different public QMC model cannot isolate solver error. Accordingly its quantitative detectability forecasts are neither used nor validated in this study. No matched numerical discrepancy has been established, and the need for an erratum cannot be assessed without the original inputs and a matched check. Detecting a mode is distinct from identifying the dark-sector origin of that mode. To the extent that a dark-sector detectability forecast, including the interpretation of Ref.~\citet{HusainOsc}, treats $f$ as composition information independent of radius, it warrants re-examination with joint structural and oscillation uncertainties. The present result does not invalidate a forecast for detecting a gravitational-wave mode itself.

\subsection*{Conclusions}
Within the jointly screened phenomenological populations studied here:
\begin{enumerate}
\item Classification of the generating family has limited generalization to held-out EoSs: 63\% without measurement error and 36\% with illustrative errors, against 25\% chance. Combining three stars sharing one EoS raises four-family accuracy to 42\%, partly through the self-bound family, but the nuclear-versus-dark-matter test remains consistent with chance (50\%; conditional EoS-cluster interval 46--55\%).
\item The median paired dark-sector shift at $1.4\,M_\odot$ is 0.50\% in $f$ (0.33--1.65\% across interaction bins), about one fifteenth of the 7.35\% nuclear spread. The corresponding median shifts in $R$ and $\Lambda$ are $-0.35\%$ and $-2.30\%$.
\item The frequency shift closely tracks the radius change, and the tidal shift is consistent with the strong compactness dependence of $\Lambda$. The paired frequency residual about the nuclear $f$--$R$ relation is 0.069\% [0.037, 0.104], against nuclear scatter of 1.73\% [1.19, 2.08] (conditional 95\% intervals). This behaviour is plausible for an effective one-fluid barotrope; independent dark and baryonic motion in a genuine two-fluid mode could break it and is not tested here. The result limits the interpretation of $f$ as an independent marker of the specified dark component, not its utility for discriminating other generating families (Table~\ref{tab:evidence}). To the extent that a dark-sector detectability forecast treats $f$ as composition information independent of radius, including the interpretation of \citet{HusainOsc}, it warrants re-examination with joint structural and oscillation uncertainties. This does not invalidate detecting a gravitational-wave mode itself.
\item In observable-shift injections, three-star discrimination first exceeds chance under the simultaneous 95\% lower-bound criterion at the tested multiplier $a=3$ (about $+1.5\%$ in $f$ and $-1.0\%$ in $R$) for both precise and low-noise error sets with all training references. At 50 and 75\% reference fractions the precise crossing stays at $a=3$, but the low-noise crossing moves to $a=2$ when averaging over repeated subsets. The crossing therefore depends on the finite-reference estimator and its uncertainty, not only measurement precision. The unamplified signal is unresolved throughout; the present data do not separate the effects of nuclear spread and reference density. This is a conditional operating point, not a detection threshold for real stars.
\item Restricted-catalogue reruns by maximum mass, interaction strength and nuclear-prior width are inconclusive after label-permutation and multiple-comparison controls; the reference sets become small.
\end{enumerate}
These conclusions are conditional on the phenomenological families, interaction prior, effective one-fluid response, independent Gaussian errors and finite reference catalogue. We do not estimate dark abundance or infer microscopic composition. Resolving the remaining questions requires new EoS draws within each interaction bin or a validated continuous-prior emulator, a larger independently drawn reference catalogue to assess convergence beyond the subsampling control, and genuine two-fluid oscillation calculations. This motivates model-aware multimessenger inference rather than reading classification accuracy as a direct composition measurement.
\section*{Data and code availability}
The complete numerical inputs, paired nuclear baselines, stellar solver, final diagnostic scripts and outputs are archived in the public GitHub repository at commit \href{https://github.com/wasif553/ML-NeutronStar-Classification/tree/9c4e28569f94ed7ae03c20ccc52709895160ad39/pdu-submission-v6/reproducibility}{\nolinkurl{9c4e28569f94ed7ae03c20ccc52709895160ad39}} (directory \texttt{pdu-submission-v6/reproducibility}). The snapshot includes the low-noise injections, reference-density control, omnibus permutation statistic, dependencies, reproduction instructions and SHA-256 checksums. The accompanying electronic supplementary archive contains the same files. The frozen input catalogue and earlier analyses are preserved at commit \href{https://github.com/wasif553/ML-NeutronStar-Classification/tree/67d6b378fb7dc10314d1dcd6dabdfc4e91def45b}{\nolinkurl{67d6b378fb7dc10314d1dcd6dabdfc4e91def45b}}, and the original population calculation at commit \href{https://github.com/wasif553/ML-NeutronStar-Classification/tree/8b955ca124dc0a44e54819278c205200d716cae2}{\nolinkurl{8b955ca124dc0a44e54819278c205200d716cae2}}. Existing copyright and applicable third-party terms remain in force; no separate dataset DOI or new software licence is asserted.
\section*{Relation to earlier work}
The earlier arXiv version (2512.23758) is superseded in its interpretation of the classification result. The published study of selected QMC stellar models \citep{HusainOsc} addresses a different question; this manuscript does not reproduce its figures or claim its detectability calculations as new.
\appendix
\section{Correction of the fixed-sequence benchmark}\label{sec:legacy}
The earlier 770-model benchmark used one EoS sequence per generating label and random row-level holdout. The reported 97.4\% test accuracy is reproducible under that protocol but is inflated by interpolation across near-duplicate stellar models. The MLP accuracy falls from approximately 97\% to 94\% and 69\% when blocks of six and 24 models are held out. Class-dependent redshift inconsistencies also exist; the redshift is excluded from the population calculations. The original data and audit are preserved in the repository for transparency. These observations do not validate or invalidate the oscillation forecasts of the separately published paper. No earlier fixed-sequence result is used as evidence of generalization to unseen EoSs in this manuscript.
\section{Restricted-catalogue statistical controls}\label{sec:controls}
The unrestricted binary control recovers 50.49\% [46.43,54.77], using the same interval convention as Table~\ref{tab:evidence}. The low-$G$ bin gives 44.95\% with all references and 44.90\% after equalizing counts. We test the conditional random-label null with 199 permutations of the selected EoS labels (one permutation per EoS, shared across repetitions). For each permutation we re-stratify the full catalogue using the original split algorithm, hold observational seeds fixed, and repeat the 20 reference subsamples per fold. Cached label-independent likelihoods give the same predictions as recomputing them; no held-out EoS enters its reference set. The null is conditional on the selected observable catalogue and does not represent a different physical interaction model. The low-$G$ null has mean 49.70\% and central range 44.13--55.93\%; the lower-tail Monte Carlo probability is 0.045 (with the plus-one correction). The smallest lower-tail probability among the nine restrictions is 0.015 for $w=0.85$, or 0.135 after a conservative Bonferroni correction for nine exploratory comparisons. As a complementary joint test of the nine conditional nulls, order the marginal permutation probabilities $p_{(i)}$ and calculate $T=\min_{1\le i\le9}(9p_{(i)}/i)$. Here $T=0.135$; the harmonic correction $H_9=\sum_{i=1}^9 i^{-1}$ gives $p_{\rm joint}=\min(1,H_9T)=0.382$ \citep{Hommel83}. This combination is conservative under arbitrary dependence of valid marginal probabilities; the overlapping, label-dependent selections are not a single exchangeable catalogue. Neither joint test establishes anomalous below-chance discrimination. Thus count imbalance alone does not explain the low-$G$ result, but the conditional null supplies no compelling evidence of a systematic reversal. The finite reference size and selection dependence still preclude a physical conclusion. The narrowest prior leaves 14 nuclear and 13 admixed EoSs.

The mass cut retains 43 nuclear, 10 self-bound, 53 admixed and 59 hyperonic-like EoSs. Under this cut the four-family precision/recall (\%) is nucleonic 33.9/39.5, self-bound 51.0/70.8, admixed 28.0/14.2, hyperonic-like 43.4/40.7. With so few self-bound EoSs, four-family changes are not interpreted as robust. After excluding CSS, the one-MR and three-star accuracies are 35.24 [32.70, 37.88]\% and 41.36 [38.32, 44.52]\% on the full sample; after the mass cut they are 36.03 [33.96, 38.21]\% and 39.59 [36.85, 42.48]\% (one-third baseline). These tests separate the contribution of the self-bound family from the binary question, without claiming continuous-prior convergence.

\end{document}